\documentclass[prb,aps,twocolumn,dcolumn,superscriptaddress,showpacs,amsfonts,amsmath,amssymb,showkeys,floatfix]{revtex4}
\usepackage{bm}
\usepackage{graphicx}% Include figure files
\begin{document}

\title{Phases of anisotropic dipolar antiferromagnets}
\date{\today}
\author{Julio F. Fern\'andez}
\affiliation{ICMA, CSIC and Universidad de Zaragoza, 50009-Zaragoza, Spain}
%\altaffiliation{IVIC}
\email[E-mail address: ] {Jefe@Unizar.Es}
\homepage[ URL: ] {http://Pipe.Unizar.Es/~jff}
\author{Juan J. Alonso}
\affiliation{F\'{\i}sica Aplicada I, Universidad de M\'alaga,
29071-M\'alaga, Spain}
\email[E-mail address: ] {jjalonso@Uma.Es}

\date{\today}
%\pacs{75.45.+j, 75.50.Xx}
%\keywords{spin reorientation, films, anisotropy, Monte Carlo}

\begin{abstract}
We study systems of classical
magnetic dipoles on simple cubic lattices with
dipolar and antiferromagnetic exchange interactions.
By analysis and Monte Carlo (MC) simulations,
we find how the antiferromagnetic phases vary with
uniaxial and fourfold
anisotropy constants, $C$ and $D$, as well as with exchange strength $J$.
We pay special attention to the spin reorientation (SR) phase,
and exhibit in detail the nature of its broken symmetries.
By mean field theory and by MC, we also obtain the
ratio of the higher ordering temperature
to the SR transition temperature, and show that 
it depends mainly on $D/C$, and
rather weakly on $J$. We find a {\it reverse} SR
transition.
\end{abstract}

\maketitle

\section {Introduction} 

Long-range magnetic order brought about by
purely dipolar interactions is somewhat rare in nature.\cite{dejongh}
Interest in the subject is
nevertheless growing.
Some of it comes from the availability of synthesized
crystals of organometallic molecules\cite{gato} that
behave at low temperature as single
spins.\cite{libro1,libro2,libro4,libro6}
Owing to the large organic mass enveloping the magnetic
cores of these molecules, dipole-dipole interactions
are then dominant.\cite{domi} Long range order has already been
observed experimentally.\cite{SMME1,SMME2,SMME3,SMME4,SMME5}
Because uniaxial anisotropy is very large,
Ising spins with dipolar interactions are reasonable models
for these systems.\cite{yo} Early rigorous work by Luttinger
and Tisza\cite{Tisza} established which type of
magnetic order obtains at low
temperature $T$ in dipolar Ising models
in each of the cubic lattices. 
The same results have been arrived at more recently by
simpler methods.\cite{nos}

No Ising model can however account for
some of the interesting collective behavior, such
as spin reorientation\cite{firstE,white,FO1,FO15,FO2} (SR)
and canted (i.e.,
noncollinear) spin configurations, that can 
be induced by purely dipolar interactions.
Accordingly, three component spin models that
include exchange as well as dipolar interactions
are often used when modeling these effects.\cite{rev2D,01CAN}
To explore the transition between the dipolar- and
exchange-dominated antiferromagnetic phases
is one of the aims of this paper.

The SR transition deserves special attention.
Thermally driven SR transitions take place
well within an ordered magnetic phase.
All spins rotate as a whole as the temperature
decreases below some SR transition temperature $T_r$.
SR transitions can be continuous, as the ones
first observed in the bulk
\cite{firstE,white}
or first order, as often observed in films.\cite{FO1,FO15,FO2}
It is important to realize that a SR {\it phase}
comes with continuous transitions.\cite{LL}
This phase
is defined by its own unique set of broken symmetries.
One of the aims of this paper is to exhibit this in detail.

It has long been realized that higher order anisotropies,
are required for the existence
of the SR phase.\cite{white,LL,earlyth} Accordingly,
we choose to study the model
Hamiltonian 
\begin{equation}
{\cal H}={\cal H}_J+{\cal H}_d+{\cal H}_A,
\label{ham}
\end{equation}
where ${\cal H}_J$ and ${\cal H}_d$ are for all exchange
and dipolar interactions, respectively,
\begin{equation}
{\cal H}_A=-D\sum_i (S_i^z)^2-C\sum_i [(S_i^x)^4+(S_i^y)^4],
\label{A}
\end{equation}
$D$ and $C$ are constants, and $S_i^\alpha$
is the $\alpha$ component ($\alpha =x,y,z$) of a unit spin $S$
at site $i$.

\begin{figure}[!ht]
\includegraphics*[width=80mm]{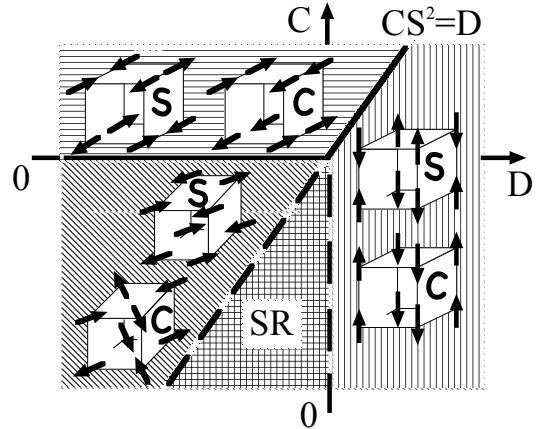}
\caption{Phase diagram for $T=0$.
The type of order labeled
$\mathfrak c$ and $\mathfrak s$
is for $-1.34\varepsilon_d\lesssim J<0$
and  $J\lesssim -1.34\varepsilon_d$, respectively.
Full and dashed thick lines stand for first--
and second--order transitions, respectively.}
\label{pattern2}
\end{figure}

The fourfold anisotropy in Eq. (\ref{A}) contributes
the desired competition to the uniaxial anisotropy.
Such a term arises naturally in tetragonal crystal lattices,
and will be seen to bring
about the canted spin phase that is shown
in the lower left quadrant of Fig. \ref{pattern2}.

To continue with the definition of the model, we let,
\begin{equation}
{\cal H}_J=-J\sum_{\langle ij \rangle}{\bf S}_i\cdot {\bf S}_j,
\end{equation}
where $\sum_{\langle ij \rangle}$ is a sum over all
nearesi neighbor {\it bonds}, and ${\bf S}_i$ is a 
classical 3-component unit spin at lattice site $i$. 
We also consider nearest neighbor exchange interactions in order to establish
how small $\mid J\mid$ must be for dipolar magnetic order to obtain.

Finally, ${\cal H}_d=(1/2)\sum_{ij}\sum_{\alpha\beta}
T_{ij}^{\alpha\beta}S_i^\alpha S_j^\beta$, where the sum is here over all
sites $i$ and $j$ except $i=j$,
\begin{equation}
T_{ij}^{\alpha\beta}=\varepsilon_d
\left(\frac{a}{r_{ij}}\right)^3\left(\delta_{\alpha\beta}-3
\frac{r_{ij}^\alpha r_{ij}^\beta}{r_{ij}^2}\right)
\label{dipene}
\end{equation}
and ${\bf r}_{ij}$ is the displacement between sites $i$ and $j$.
We let $E_A$, $E_J$, and $E_d$ be the values of
${\cal H}_A$, ${\cal H}_J$, and ${\cal H}_d$ for whatever
state is specified.

Since we treat all spins as classical unit vectors, no
quantum effect is taken into account. Accordingly, we
disregard hyperfine
interactions, since they have no effect on the thermal equilibrium behavior
of electronic spins when the latter are treated classically.\cite{hyperf}
Quantum effects, such as the enhancement of the 
transverse field that is necessary in order to destroy long-range 
order in LiHoF$_4$ below approximately 1 K,\cite{aeppli2}
are therefore beyond the scope of this paper. Thus, some numbers we obtain,
such as the crossover value of J, below which dipolar interactions
become dominant, may change some when hyperfine interactions are included
in a quantum treatment. On the other hand, our main results,
the insensitivity to the value of $J$ to the ratio of the ordering
temperature to the SR temperature, as well as the existence of a reverse
SR transition, are of a rather qualitative nature, and are
therefore expected to be insensitive to quantum effects.

Simple cubic (sc) lattices and zero
applied magnetic field $H$ are assumed throughout.
We only work with
$L\times L\times L$
box like systems, and let dipole-dipole interactions act between
each spin and all other spins within
an $L\times L\times L$ box centered on it.
We use periodic boundary conditions, because they
give faster convergence
towards the $L\rightarrow \infty $ limit than
free boundary conditions do.
In addition, because we work with antiferromagnets, and not
ferromagnets, convergence is faster than would 
otherwise be.
Most of our results follow from simulations
for $L=8$ and $L=16$. This
would be insufficient for critical behavior work
but is adequate for our purposes, that is,
establishing which type of magnetic order
obtains and the corresponding phase boundaries.
It is worth recalling that 
thermal equilibrium results
obtained for $H=0$ for large cubic-shaped systems
can, by virtue of Griffith's theorem,\cite{griff}
be generalized to other shapes in three dimensions (3D).

Our simulations follow the standard Metropolis
Monte Carlo algorithm.\cite{metrop} More specifically, we start
simulations with an initial configuration
in which all spins point in either random or 
parallel directions. We next compute
the dipolar field at each site. Time evolution
takes place as follows. A spin is chosen at random
and temporarily pointed in a new random direction. Let $\Delta E$
be the corresponding
energy change. If $\Delta E<0$, the temporary direction becomes
permanent. If on the other hand $\Delta E>0$ the temporary
direction becomes permanent only with probability
$\exp (-\Delta E/k_BT)$, where $T$ is the system's temperature.
The field changes that ensue at every site in the system if
the new spin direction is accepted is then computed, thus
updating the field values everywhere on the system.

The plan of the paper and a list of the results obtained follow.
In Sec. \ref{gs}, we study all the phases,
except the SR phase (see Fig. \ref{pattern2}), for
all $J\leq 0$. To this end we first define a canted state
which lets all spins point along the easy magnetization axes
for any $C$
and $D$ while
$E_d$ remains invariant if the system's shape
is cubic. This is the main device which enables us to conclude
in Sec. \ref{gs} that $\mathfrak c$
and $\mathfrak s$, exhibited in Fig. \ref{pattern2},
are shown to be the ground state configurations 
for $-1.34\varepsilon_d<J<0$ and $J<-1.34\varepsilon_d$,
respectively. We also report MC data which suggests that,
as in the Landau theory,\cite{LL}
the $xy$-collinear to $xy$-canted transition line
is first order. We also report results from MC simulations
for temperature driven phase transitions
between paramagnetic and $xy$-collinear and
between $xy$-collinear and $xy$-canted phases.
In Sec. \ref{SR} we study
the SR phase.
In an anisotropy diagram we show
the free energy minima that follow from MC
simulations for the SR phase. The broken
symmetries are clearly exhibited. 
This suggests
how to write mean field equations for the SR transition,
from which we obtain $T_r$ as well as the higher ordering
temperature as a funtion of $D/C$.
We show that the ratio of these two temperatures
depends mainly on the $D/C$ ratio and rather weakly on $J$,
and mean field
yields approximately the same value for it as MC simulations do.
For $D<0$ and $0.8\lesssim D/C<1$, cooling through the paramagnetic
phase first brings the system into $xy$-canted ordering. The
SR phase is encountered at a lower temperature. For $D<0$ but
$0<D/C\lesssim 0.8$, cooling through the paramagnetic
phase first brings the system into $z$-collinear ordering,
and the SR phase is encountered at a lower temperature.
An interesting phase diagram obtains at $D/C\sim 0.8$,
and a reverse spin reorientation
that is temperature driven is observed. A complete
spin reorientation, from the easy magnetization
axis into the perpendicular plane, first takes place upon cooling
below the paramagnetic phase, followed, upon further
cooling, by a reverse
spin reorientation towards the easy magnetization axis.
This is so for
$J=0$ as well as for $\varepsilon_d=0$.

\section{collinear and canted phases}
\label{gs}
 
In this section we study all the phases shown in
Fig. \ref{pattern2}, except the SR phase. 
We first study how the ground state of a purely
dipolar system varies with $D$ and $C$.
We then extend this to all $J<0$, and
finally consider nonzero temperatures.

\subsection{The canted state}
\label{cant}

In this subsection we define the canted state $\mathfrak c$.
We do this in order to avoid the difficulty which follows from the
the fact that dipolar interactions are not rotationally
invariant, thus making the task of
simultaneously minimizing $E_A$
and $E_d$ nontrivial.
The canted state $\mathfrak c$ enables one to do this
minimization. Let
\begin{equation}
S_i^z=\tau_i^{z}\cos \theta,\; S_i^y=\tau_i^{y}\sin \theta \sin\phi,\;
S_i^x=\tau_i^{x}\sin \theta \cos \phi,
\label{ctx}
\end{equation}
where $\theta$ is the angle between the spin
vector and the $z$-axis, $\phi$ is the azimuthal angle,
$\bm{\tau}_i\equiv [\tau_i^x,\tau_i^y, \tau_i^z]$ is given by
\begin{equation}
\bm{\tau}_i=[(-1)^{y(i)+z(i)},
(-1)^{x(i)+z(i)},
(-1)^{x(i)+y(i)}],
\label{taud}
\end{equation}
$[x(i), y(i), z(i)]$ is the position of site $i$
and $x(i)$, $y(i)$, and $z(i)$ are all integers.

Before proceeding with the argument, we point out
some basic features of the canted state we
have just defined. Note how ${\bf \tau}_i$ varies
with $i$. A canted spin configuration therefore
follows from Eqs. (\ref{ctx}) and (\ref{taud})
if $\theta \neq 0$ and $\theta\neq\pi /2$. For $\theta =\pi /2$,
a canted spin configuration also obtains if
$\phi \neq 0$ and $\phi\neq\pi /2$, as, for instance,
the ${\mathfrak c}$ state in the lower left hand
quadrant of Fig. \ref{pattern2}. The fact that our definition
of a $\mathfrak c$ state gives a collinear spin
configuration for these few special cases (see the ${\mathfrak c}$
state in the lower right hand corner of Fig. \ref{pattern2}
for $\theta =0$)
should not be too confusing.

To start our argument, note that $\theta =0$ gives the minimum
value of $E_d$ for sc lattices in the bulk.
\cite{Tisza,griff} It is next
shown that the dipolar energy $E_d$
of cubic-shaped systems is independent
of $\theta$ and of $\phi$ in the canted state.\cite{Rinv}
In order to see this, consider first
$\textbf {S}_i\cdot {\bf S}_j$, which upon substitution
of Eq. (\ref{ctx}) becomes
$\tau_{i}^{z}\tau_{j}^{z}\cos^2\theta+
\tau_{i}^{y}\tau_{j}^{y}\sin^2\theta\sin^2\phi+
\tau_{i}^{x}\tau_{j}^{x}\sin^2\theta\cos^2\phi$.
Now, it is easy to check, that the {\it sum}
$\sum_{i,j}{\bf S}_i\cdot {\bf S}_jf(r_{ij})$, where
$f(r_{ij})$ is any function of $r_{ij}$, over a sc
lattice bounded by the surface of a cube
is independent of $\theta$ and of $\phi$,
since $\sum_{i,j}\tau_{i}^{z}\tau_{j}^{z}f(r_{ij})
=\sum_{ij}\tau_{i}^{y}\tau_{j}^{y}f(r_{ij})
=\sum_{ij}\tau_{i}^{x}\tau_{j}^{x}f(r_{ij})$
then. Thus, the first of the two terms of the dipolar interaction
is independent of $\theta$ and of $\phi$, and, by the way, so is the
exchange energy.
Now, consider $w_{ij}\equiv
(x_{ij}S_i^x+y_{ij}S_i^y+z_{ij}S_i^z)
(x_{ij}S_j^x+y_{ij}S_j^y+z_{ij}S_j^z)$, from the second term of
the dipolar interaction. Again, $\sum_{ij}w_{ij}/r_{ij}^3$ over
a sc lattice bounded by the surface of a cube is independent of
$\theta$ and of $\phi$, since (1) cross terms do not contribute
to the sum, by reflection symmetry, and (2)
$\sum_{i,j}\tau_{i}^{z}\tau_{j}^{z}/r_{ij}^3
=\sum_{ij}\tau_{i}^{y}\tau_{j}^{y}/r_{ij}^3=
\sum_{ij}\tau_{i}^{x}\tau_{j}^{x}/r_{ij}^3$.
Therefore, $\theta$ and $\phi$ can be freely chosen in Eq.
(\ref{ctx}) in order to
minimize $E_A$, at no cost to $E_d$,
which is the desired result for bulk systems.

Finally, since
the minimum value of $E_A$
can be reached by
appropriate choice of
$\theta$ and of $\phi$ in Eq.
(\ref{ctx}), it follows that minimization of $E_A$ with
the $\mathfrak c$ state gives the ground state.

\begin{figure}[!ht]
\includegraphics*[width=80mm]{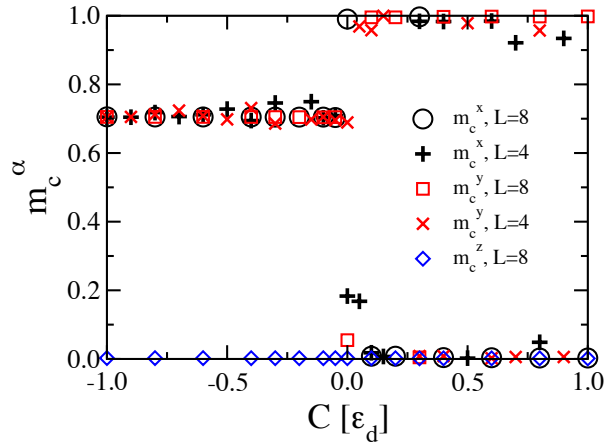}
\caption{(Color online) Order parameter $m_c^\alpha$, for
$\alpha =x, y, z$,  vs $C$
for systems of $L\times L\times L$ dipoles
on sc lattices at $T=0.04\varepsilon_d/k_B$,
where $k_B$ is Boltzmann's constant, for $D=-3\varepsilon_d$.
A transition between the $xy$-collinear and
$xy$-canted phases is clearly exhibited at $C=0$.
In order to reach equilibrium, $T$
was lowered from $T=2\varepsilon_d/k_B$
in $\Delta T=-0.02\varepsilon_d/k_B$ steps.
$2 \times 10^4$ MC sweeps were made at each temperature step.
Equilibrium values come from averages
over $2 \times 10^5$ MC sweeps.}
\label{CantCol}
\end{figure}

The phases obtained for the ground state
by this procedure if $J=0$ are shown in Fig. \ref{pattern2}.
We refer to the states on the right hand side, upper left
hand side and lower left hand sides of the diagram as
$z$-collinear, $xy$-collinear, and $xy$-canted phases.
The nature of the SR phase is the subject of Sec. \ref{SR}.

Note that the $xy$-collinear phase is
unstable with respect to any nonvanishingly small $C<0$.
[On the other hand (see below), the $xy$-canted phase obtains
in finite nonzero temperature only for a sufficiently
small $C<0$.]

Before we proceed any further,
we need some additional definitions. Let
\begin{equation} 
m_{\mathfrak c}^\alpha =N^{-1}\sum_i  S_i^\alpha \tau^{\alpha}_i.
\end{equation}
In a $\mathfrak c$
state, $m_{\mathfrak c}^z=\cos \theta$,
$m_{\mathfrak c}^x=\sin \theta\cos \phi$, and
$m_{\mathfrak c}^y=\sin \theta\sin \phi$.

According to Landau's theory,
the transition between the $z$-collinear and the $xy$-plane
phases, as well as the transition between the $xy$-collinear
and the $xy$-canted phases,
is of first order.\cite{LL} This has been observed for
the $z$-collinear to $xy$-plane transition
in films.\cite{FO15,01CAN,nueva}
The results we have obtained from MC simulations
also support this conclusion.

On the other hand, we know of no experimental or MC evidence
about the nature of the $xy$-collinear to $xy$-canted transition.
Data points from MC simulations we have performed are
plotted in Fig. \ref{CantCol}.
Note: (1) $m_{\mathfrak c}^z=0$ for all $-1<C/\varepsilon_d<1$,
(2) $m_{\mathfrak c}^x=m_{\mathfrak c}^y$
for all $C/\varepsilon_d<0$, and (3)
either (a) $m_{\mathfrak c}^x=0$ and $m_{\mathfrak c}^y=1$
or (b) $m_{\mathfrak c}^x=1$ and $m_{\mathfrak c}^y=0$
for $C>0$.
A first order phase transition, between two in plane
spin configurations, one along the $x$ or $y$ axes and
the other one with $\phi =\pi /2$, is clearly suggeted.

\subsection{Exchange interaction}

The effect exchange interactions
have on the ground state is studied in this subsection.

Let's first define the collinear $\mathfrak s$ state,
illustrated in Fig. \ref{pattern2} by one of
the two sets of the states.
In this state, all
nearest neighbor spins to spin ${\bf S}$ point
opposite to ${\bf S}$ (as in Fig. \ref{pattern2}).
Alternatively,
in a $\mathfrak s$ state,
\begin{equation}
S_i^z=\eta_i\cos \theta,\; S_i^y=\eta_i\sin \theta \sin\phi,\;
S_i^x=\eta_i\sin \theta \cos \phi,
\label{cex}
\end{equation}
where
\begin{equation}
\eta_i\equiv (-1)^{x(i)+y(i)+z(i)}
\end{equation}

It makes sense to also define
\begin{equation}
 m_{\mathfrak s}^\alpha =N^{-1}\sum_i  S_i^\alpha \eta_i.
\end{equation}
In a $\mathfrak s$ state,
$m_{\mathfrak s}^z=\cos \theta$,
$m_{\mathfrak s}^x=\sin \theta\cos \phi$, and
$m_{\mathfrak s}^y=\sin \theta\sin \phi$, but in a $\mathfrak c$ state,
${\bf m}_{\mathfrak s}=0$.

We showed in Sec. \ref{cant} that $\mathfrak c$
is the ground state if $J=0$ and that $E_d$
is independent of $\theta$ and
$\phi$ in Eq. (\ref{ctx}). In
addition, $E_d=-2.67\varepsilon_d$ then.\cite{nos}
Note also that Eqs. (\ref{ctx}) and (\ref{taud}) imply
$E_J=J$ for all $\theta$ and $\phi$ in a $\mathfrak c$
state. Similarly,
it can be easily checked that
in a $\mathfrak s$ state, $E_d=0$, independently of the direction
of ${\bf S}$, and that, $E_J=-3J$.

Therefore, of the above two states, the $\mathfrak c$ 
state gives the lower energy in the $-1.34\varepsilon_d<J<0$
range and the $\mathfrak s$ state gives the lower
energy in the $J<-1.34\varepsilon_d$ range. 
Monte Carlo simulations show that no other state
gives a lower energy in the whole $J<0$ range, that is,
\begin{equation}
E=-2.67\varepsilon_d+J+E_A
\end{equation}
if
$ -1.34\varepsilon_d\lesssim J<0$ and 
\begin{equation}
E=3J+E_A
\end{equation}
if $J<-1.34\varepsilon_d$.

Clearly, $\partial E/\partial J$ is discontinuous
at $J\simeq -1.34 \varepsilon_d$, implying, a first order transition
betwwen purely dipolar induced canted 
(i.e., a $\mathfrak c$ state) and collinear antiferromagnetic
phases (i.e., a $\mathfrak s$ state) for all $C$ and $D$.

\subsection{$T\neq 0$}

\begin{figure}[!ht]
\includegraphics*[width=80mm]{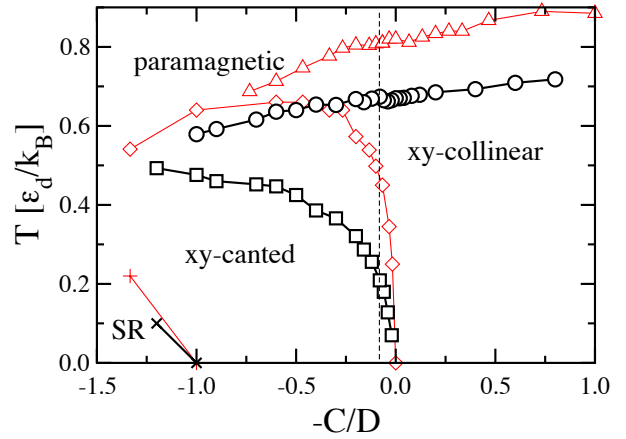}
\caption{(Color online) Transition temperatures and phases
for systems with only dipolar interactions ($J=0$),
and two values of $D$: $-3\varepsilon_d$ ($\triangle$ and
$\lozenge$), and
$-0.5\varepsilon_d$ ($\circ$ and $\Box$).
$+$ are for the SR phase boundary.
The data come from MC simulations of systems
of $8\times 8\times 8$ spins on sc lattices.
The vertical dashed line stands for the cooling
path that was taken to obtain the data points shown in
Fig. \ref{ree2}.}
\label{ree1}
\end{figure}

\begin{figure}[!ht]
\includegraphics*[width=80mm]{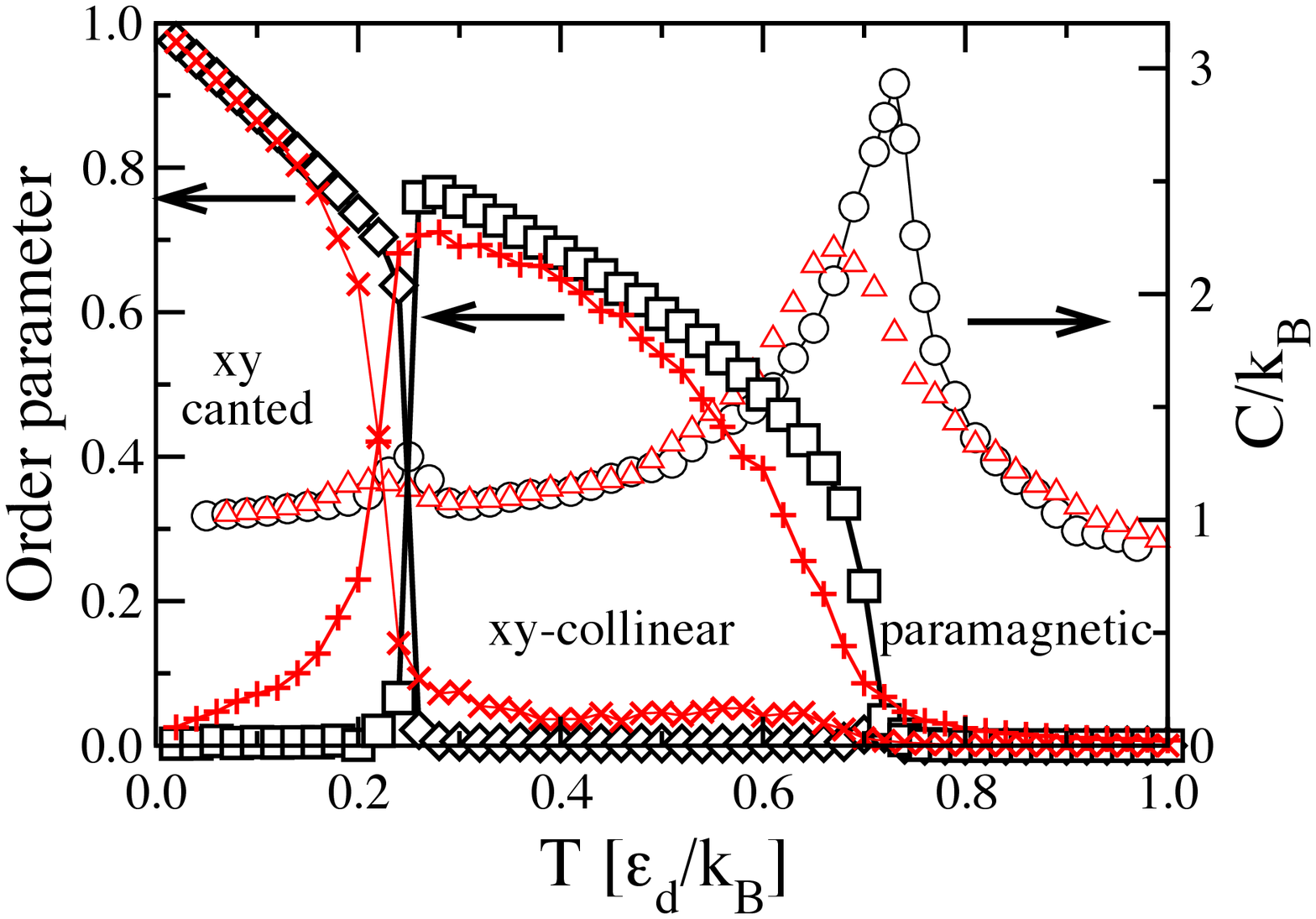}
\caption{(Color online) $\mid [m_c^x]^2-[m_c^y]^2 \mid$
$[2 m_c^x m_c^y ]^2$ and $C/k_B$ 
vs $T$ for systems of $L\times L\times L$
spins with $J=0$, and $D=-0.5 \varepsilon_d$,
and $C=0.05 D$.
For $L=8$, $+$ (red online), $\times$
(red online), and $\triangle$ (red online)
stand for $\mid [m_c^x]^2-[m_c^y]^2 \mid$,
$[2 m_c^x m_c^y ]^2$ and $C/k_B$, respectively;
for $L=16$, $\Box$, $\lozenge$, and $\circ$
stand for $\mid [m_c^x]^2-[m_c^y]^2 \mid$,
$[2 m_c^x m_c^y ]^2$ and $C/k_B$, respectively.
For $L=16$ ($L=8$), all data points
stand for averages over $10^5$
($2.5 \times 10^5$) MC sweeps.}
\label{ree2}
\end{figure}

Of all thermally driven
transitions between any two of the paramagnetic,
$xy$-collinear, and $xy$-canted phases
(the SR phase is treated in Sec. \ref{SR}), the most
interesting one is the one between
the $xy$-collinear and the $xy$-canted phases. It
is illustrated in Figs. \ref{ree1} and \ref{ree2}.
This effect has been discussed before for thin films.\cite{thre}
Note how the variation of
$\mid [m_{\mathfrak c}^x]^2-[m_{\mathfrak c}^y]^2\mid$
and of $\mid m_{\mathfrak c}^x m_{\mathfrak c}^y \mid$
at $T\simeq 0.24\varepsilon_d$
becomes sharper as the systems size increses. This
suggests a first order phase transition, in accordance
with Landau's theory.\cite{LL}

\section{The spin reorientation phase}
\label{SR}

\subsection{$T=0$}

\begin{figure}[!ht]
\includegraphics*[width=80mm]{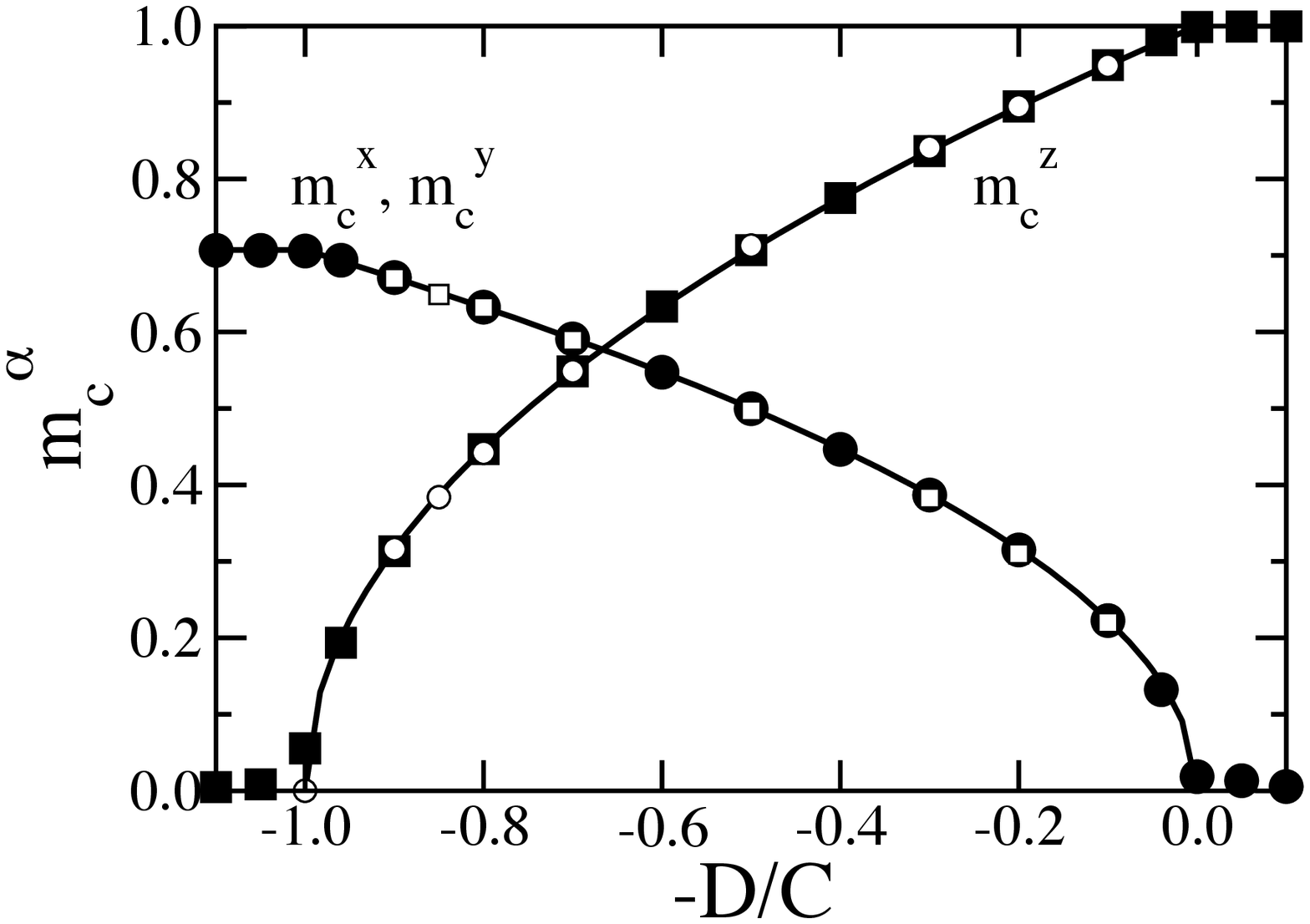}
\caption{$m_{\mathfrak c}^\alpha$
vs $-D/C$ for $\alpha =x,y,z$ and $T=0$.
$\bullet$ stands for both $m_{\mathfrak c}^x$
and $m_{\mathfrak c}^y$,
and $\blacksquare$ stands for $m_{\mathfrak c}^z$,
for $C=-0.5\varepsilon_d$ and $J=0$; $\Box$ stands
for both $m_{\mathfrak c}^x$
and $m_{\mathfrak c}^y$,
and $\circ$ stands for
$m_{\mathfrak c}^z$,
for $\varepsilon_d=0$ and $C=J<0$.
All symbols are from MC simulations in which $T$
was lowered from $T\gg \varepsilon_d,\mid J \mid$
to $T\ll \varepsilon_d, \mid J\mid$.
Continuous lines are from Eq. (\ref{Fdetb}) and below.}
\label{Sgst}
\end{figure}

We first explore SR at $T=0$.
We start by minimizing the anisotropy energy $E_A$ for
all $C$ and $D$ in
the SR phase (see Fig. \ref{pattern2}). Recall from Sec. \ref{gs}
that $E_A$ can be minimized freely for
all $J\leq 0$.
We obtain
\begin{equation} 
{\bf S}=(\pm u,\pm u, \pm v)
\label{Fdetb}
\end{equation}
for $C <D<0$, where $ u=\sqrt{D/2C}$ and $v=\sqrt{1-2u^2}$.
Thus, varying $D$ through the $C<D<0$ range
forces the spin directions to vary in the ground state from
the $z$-collinear state shown in Fig. \ref{pattern2} to the
$xy$-phase.
This is analogous to the phenomenological theory of SR.\cite{LL}

We have simulated cooling from high
temperatures down to $k_BT\ll \varepsilon_d, J$ for
various values of $C$ and $D$: (1) purely
dipolar interacting 3D systems (i.e., $J=0$);
(2) antiferromagnetic 3D systems with only
nearest neighbor interactions
(i.e., $J<0$, $\varepsilon_d=0$).
The MC results obtained as well as the numbers
that follow from Eq. (\ref{Fdetb}) are plotted in Fig. \ref{Sgst}.
The data points fall on the predicted curves,
independently of $J$.

Note that Eq. (\ref{Fdetb}) allows
8 different spin directions. This suggests which
symmetries are broken in the SR phase. We examine this
idea in more detail in the following section.

\subsection{The temperature driven transition}
\label{ttr}

\begin{figure}[!ht]
\includegraphics*[width=80mm]{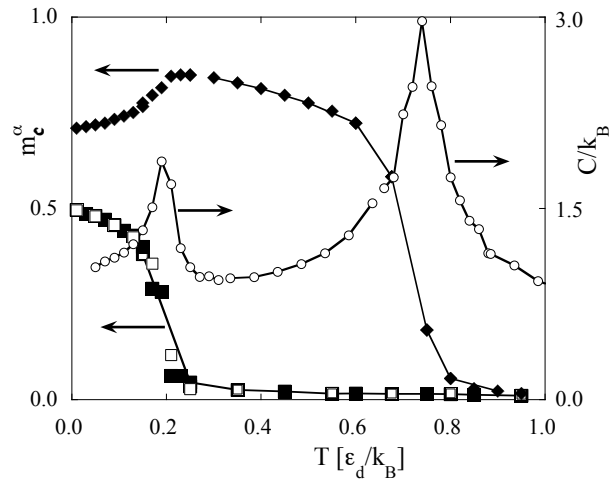}
\caption{$m_{\mathfrak c}^z$ ($\blacklozenge$),
$m_{\mathfrak c}^y$
($\blacksquare$),
$m_{\mathfrak c}^x$ ($\Box$),
and $C/k_B$ ($\circ$) vs $T$.
All data points
come from MC simulations of $8\times 8\times 8$
spins on sc lattices for $J=0$,
$C=-\varepsilon_d$, and $D=0.5C$.
At each value of $T$, at least $10^5$ MC sweeps were
made. Lines are guides to the eye.}
\label{SRtr2} 
\end{figure}

\begin{figure}[!ht]
\includegraphics*[width=80mm]{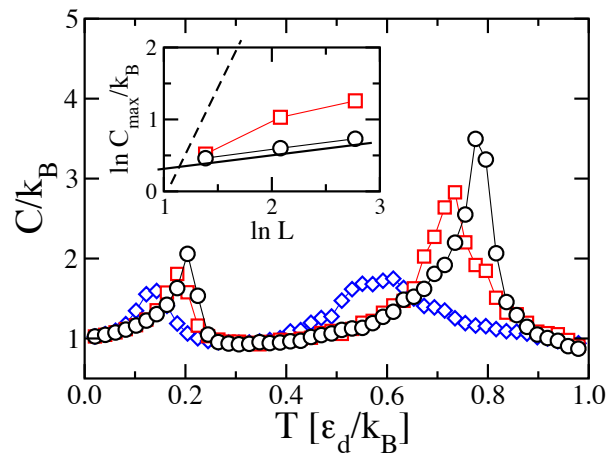}
\caption{Specific heat vs temperature for systems of
$L\times L\times L$ dipoles, with $J=0$,
$C=-\varepsilon_d$ and $D=0.5C$.
$\circ$, $\square$ and $\diamond$ stand for $L=16, 8$ and $4$
respectively. $T$ was lowered in
$\Delta T=-0.01$ steps. At each value of
$T$, we made $5\times 10^4, 5\times 10^5$ and $2 \times 10^7$ MC sweeps
for $L=16, 8$ and $4$, respectively. The high (low) temperature peak
corresponds to the paramagnetic to $z$-collinear phase ($z$-collinear to SR
phase) transition. Inset: $\log~C_{max}/k_B$ vs $\log L$
for both transitions. $\circ$ ($\square$) stands for
the high (low) $T$ peaks.
The slope of the dashed line is
for a first order transition ($C\sim N$ then).
The slope of the full line is $0.2$, in accordance
with a countinuous transition in which
$\alpha /\nu\approx 0.2$.}
\label{seconorder}
\end{figure}

\begin{figure}[!ht] 
\includegraphics*[width=80mm]{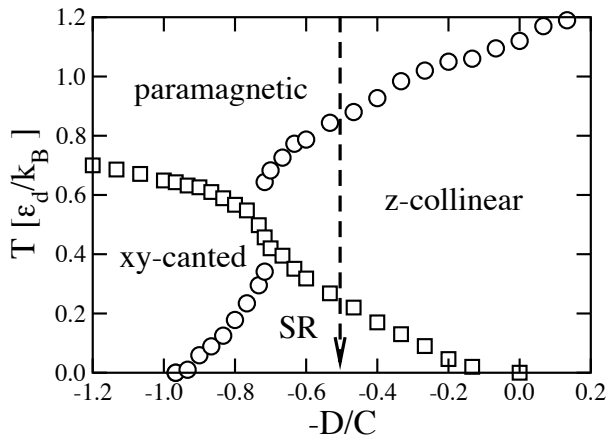}
\caption{(a) Phases of dipolar ($J=0$) antiferromagnets
for $C=-3\varepsilon_d$. All
data points come from MC simulations of
$8\times 8\times 8$ spins. The dashed line
is for the cooling path from which the data
points in Fig. \ref{SRtr2} follow.}
\label{SRph}
\end{figure}

In Fig. \ref{SRtr2}, $ m_{\mathfrak c}^\alpha$ for $\alpha =x,y,z$,
and the specific heat $C$
are plotted vs $T$ for $J=0$, and
$D=-0.5\varepsilon_d$, $C=-\varepsilon_d$.
The data points are
from MC simulated cooling. Data points obtained from heating from ordered
states, using Eq. (\ref{Fdetb}) and below, do not differ from the ones shown.

Specific heat curves obtained for
various system sizes are shown in Fig. \ref{seconorder}.
The data are not good enough
for an accurate value of any critical index, but they
suggest the transitions are continuous at both of the
two transition temperatures.

We have obtained additional specific heat and
${\bf m}_c$ curves (as the ones in Fig. \ref{SRtr2})
for $J=0$ and $C=-3\varepsilon_d$, from further MC simulations.
The data points shown in Fig. \ref{SRph} for
the two transition temperatures as well as the labels
shown for the phases  follow from such curves.

For all $J<0$, the phase diagram shown
in Fig. \ref{SRph} is qualitatively the same, except that
for $J<-1.3\varepsilon_d$, the $xy-$canted phase is
replaced by the collinear $xy$ phase shown in Fig. \ref{pattern2}
for $C,D<0$. 

\subsection{Broken symmetries}
\label{ph}

\begin{figure}[!ht]
\includegraphics*[width=80mm]{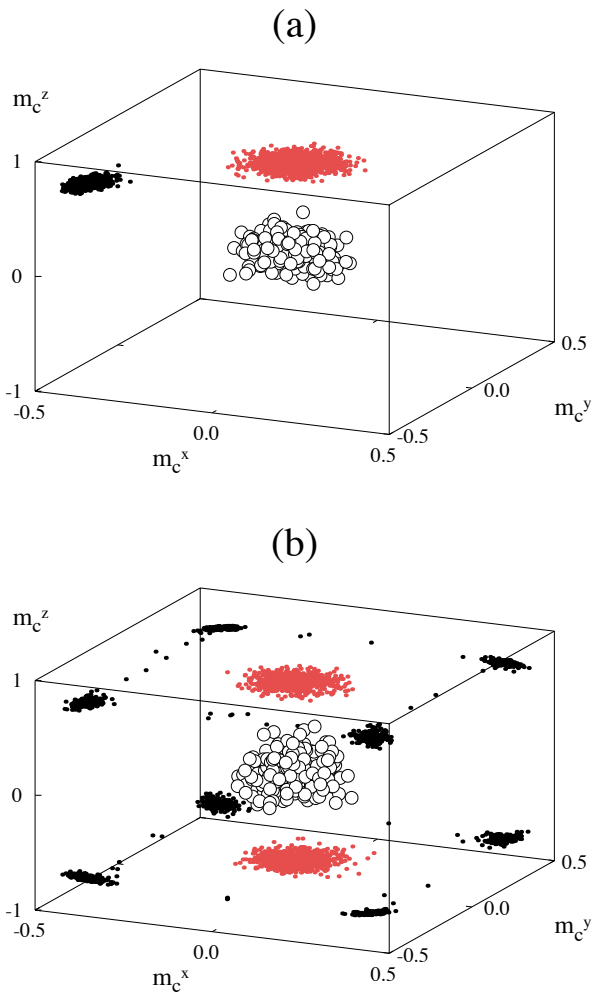}
\caption{(Color online)(a)
${\bf m}_{\mathfrak c}$ values measured
every $10^3$ equally
spaced MC sweeps in equilibrium
for systems of $8\times 8\times 8$ dipoles with
$J=0$, $C=-\varepsilon_d$ and $D=0.5C$. Each set of $\circ$,
gray (red online), and black dots stand for
$k_BT/\varepsilon_d=0.9$, $0.5$, and $0.15$, respectively.
All data points for each $T$ come from a single
MC run of $4\times 10^6$ MC sweeps each.
In each MC run, before we take any data,
we let the system equilibrate
by lowering
the temperature from $T=2\varepsilon_d/k_B$ (deep in
the paramagnetic phase) in steps of
$\Delta T=-0.1$, of $4\times 10^3$ MC sweeps each.
(b) Same as in (a) but each of the three sets of
points comes from
$10^3$ independent short MC runs (i.e., first cooling
from $T=2\varepsilon_d/k_B$ in each of the $10^3$ MC runs)
of $4\times 10^4$ MCS steps each.}
\label{BrSym}
\end{figure}

In this section we illustrate the symmetries of the various
magnetic phases a system goes through when cooling from the
paramagnetic into the SR phase.
To this effect, a statistical sample of values ${\bf m}_{\mathfrak c}$
takes up throughout time in each of the three phases
--paramagnetic, $z$-collinear, and $SR$--
that obtain for $C=-\varepsilon_d$ and $D=0.5C$ is shown
in Figs. \ref{BrSym}(a) and \ref{BrSym}(b). 

Now
\begin{equation}
p({\bf m}_{\mathfrak c})=Z^{-1}e^{-F({\bf m}_{\mathfrak c})/k_BT},
\end{equation}
where $p({\bf m}_{\mathfrak c})$ is the probability
to find a ${\bf m}_{\mathfrak c}$ value
and $F({\bf m}_{\mathfrak c})$ is the free energy.
It follows that cloud densities in Figs. \ref{BrSym}(a) and \ref{BrSym}(b)
are proportional to $\exp [-F({\bf m}_{\mathfrak c})/k_BT]$.
The clouds therefore stand for neighborhoods of the
minima of $F({\bf m}_{\mathfrak c})$ for each of the three
phases. 

The data points shown in Fig. \ref{BrSym}(a)
for each value of $T$ come from one single
MC run. This is in contrast to the procedure
used to obtain the data points shown in 
Fig. \ref{BrSym}(b), which were obtained from
$10^3$ independent MC runs 
of $4\times 10^4$ MCS steps each.

Clearly, all the symmetries that are broken in the
$xy$-canted phase, in addition to the
symmetries that are broken in the
$z$-collinear phase,
are broken in the SR phase.

\subsection{Mean field}

\begin{figure}[!ht]
\includegraphics*[width=80mm]{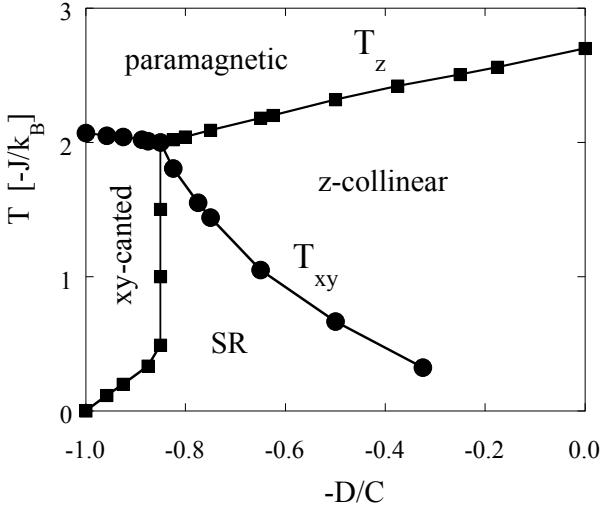}
\caption{Mean field transition temperatures $T_z$ and $T_{xy}$
and the nature of the antiferromagnetic phases
vs $-D/C$ for 3D, nearest neighbor interactions only (i.e., $\varepsilon_d=0$),
$J<0$, and $C=4J$.}
\label{MF} 
\end{figure}

It is interesting to calculate the mean field
transition
temperatures associated with the SR phase.
We assume a magnetic order as
either in a canted or a collinear state, and that only
${\bf m}_{\mathfrak c}$ and ${\bf m}_{\mathfrak s}$ remain
to be determined. More precisely, we assume
\begin{equation}
\langle S_i^\alpha\rangle=m_{\mathfrak c}^\alpha\tau_i^\alpha
+m_{\mathfrak s}^\alpha\eta_i.
\label{14}
\end{equation}
We now write the mean field equations.
First note that multiplying the above equation by $\tau_i^\alpha$
and summing over two nearest neighbor sites along the $\alpha$
direction gives,
\begin{equation}
2m^\alpha_{\mathfrak c}=\sum_{\alpha,i=1,2}
\tau_i^\alpha\langle S_i^\alpha\rangle .
\label{15}
\end{equation}
Similarly, one obtains
\begin{equation}
2m^\alpha_{\mathfrak s}=\sum_{\alpha,i=1,2}
\eta_i\langle S_i^\alpha\rangle .
\label{16}
\end{equation}
We now write the main mean field equation,
\begin{equation}
\langle{\bf S}_i\rangle =Z^{-1}\sum_{{\bf u}}
 {\bf u}\exp [-\varepsilon_i({\bf u})/k_BT]
\label{17}
\end{equation}
where the sum is over all directions of the
unit vector ${\bf u}$,
\begin{equation}
\varepsilon_i({\bf u})=-{\bf h}(i){\bf \cdot u}+\varepsilon_A({\bf u}),
\label{18}
\end{equation}
\begin{equation}
\varepsilon_A({\bf u})=-Du_z^2-C(u_x^4+u_y^4),
\end{equation}
${\bf h}(i)={\bf h}_J(i)+{\bf h}_d(i)$, 
\begin{equation}
h_J^\alpha(i)=J(m_{\mathfrak c}^\alpha\sum_{ni}\tau_i^\alpha
+m_{\mathfrak s}^\alpha\sum_{ni}\eta_i),
\label{19}
\end{equation}
the sum here is over all nearest neighbors of site $i$,
\begin{equation}
h_d^\alpha(i)=\sum_j\sum_\beta T_{ij}^{\alpha\beta}(m_{\mathfrak c}^\beta
\tau_j^\beta+m_{\mathfrak s}^\beta\eta_j),
\label{20}
\end{equation}
and, finally,
\begin{equation}
Z=\sum_{\bf u} \exp [-\varepsilon_i({\bf u})/k_BT].
\label{21}
\end{equation}

The above equations simplify near the phase boundaries.
Consider, for instance, the boundary between
the paramagnetic and the $z$-collinear phases
in Fig. \ref{MF}. Let $T_z$ define the
the boundary between a phase where $m_{\mathfrak c}^z$
and $m_{\mathfrak s}^z$ vanish and a phase where
either $m_{\mathfrak c}^z$ or $m_{\mathfrak s}^z$
does not. $T_{xy}$ is similarly defined for the
$x$ and $y$ compnents of ${\bf m}_{\mathfrak s}$
and ${\bf m}_{\mathfrak c}$.

Then, $\exp (-{\bf h\cdot u}_i/k_BT)$
can be linearized, and 
\begin{equation}
kT_z=6JZ^{-1}\int d\Omega \;u^2_ze^{-\varepsilon_A({\bf u})},
\label{Tz}
\end{equation}
where
\begin{equation}
Z=\int d\Omega \;e^{-\varepsilon_A({\bf u})},
\end{equation}
and $\int d\Omega$ is an integral over all directions of ${\bf u}$.
Similar equations obtain for the 
the paramagnetic-$xy$-canted phase boundary.
For other phase boundaries, linearization
cannot be carried out for all spin components,
whence slightly more complicated
systems of equations ensue.
We
have solved the equations for the phase
boundaries numerically.
The solutions for $\varepsilon_d=0$
and $C=4J<0$ are
shown in Fig. \ref{MF}. 

\begin{figure}[!ht]
\includegraphics*[width=80mm]{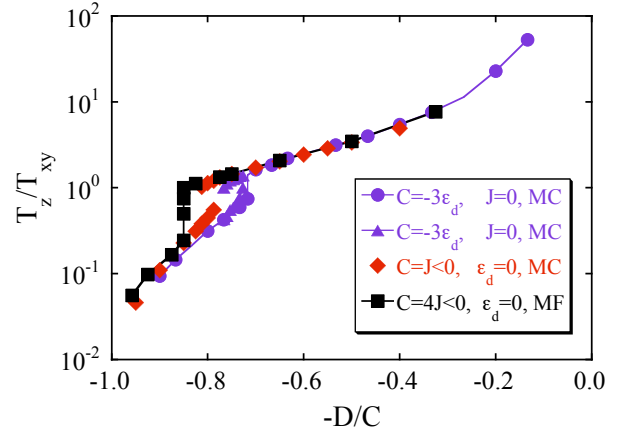}
\caption{(Color online) Temperature ratio $T_{xy}/T_z$
vs $-D/C$, for the shown values of $\varepsilon_d$,
$J$, and $C$ in cubic-shape systems.
MF stands for mean field theory results.
$\bullet$ and $\blacktriangle$ stand for $8\times 8\times 8$
and $16\times 16\times 16$ spin systems, respectively.
Data points from MC
simulations are for
$8\times 8\times 8$
spins.
Lines are guides to the eye.}
\label{Tratio}
\end{figure}

\begin{figure}[!ht]
\includegraphics*[width=80mm]{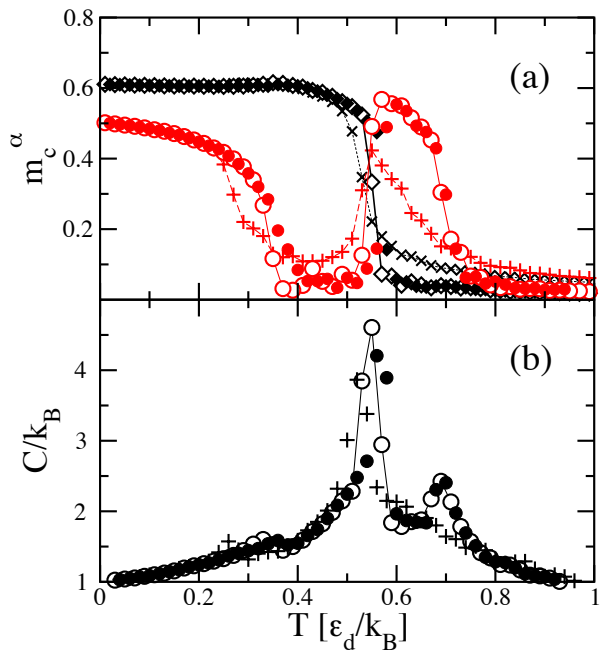}
\caption{(a) $m_{\mathfrak c}^\alpha$, for
$\alpha = y, z$ vs $T$
for systems of $L\times L\times L$ dipoles with $J=0$,
$C=-3\varepsilon_d$ and $D=0.747 C$.
Data points for $m_{\mathfrak c}^x$ would fall right
on top of data points for $m_{\mathfrak c}^y$
and are not shown.
Open (closed) symbols stand for data points
from MC simulations in which $T$ was lowered
(raised), starting from
$T=\varepsilon_d$ ($T=0.01 \varepsilon_d$).
Circles and rhombi stand for $\alpha =z$ and $x$,
respectively, both for $L=16$.
+ ($\times$) stand for $\alpha=z$ and $\alpha=y$
for systems with $L=8$.
All data points come from averages of
$\mid m_{\mathfrak s}^\alpha\mid$
over some $10^5$ MC sweeps.
(b) Same as in (a) but for $C/k_B$ vs $T$.}
\label{2SRd}
\end{figure}

\begin{figure}[!ht]
\includegraphics*[width=80mm]{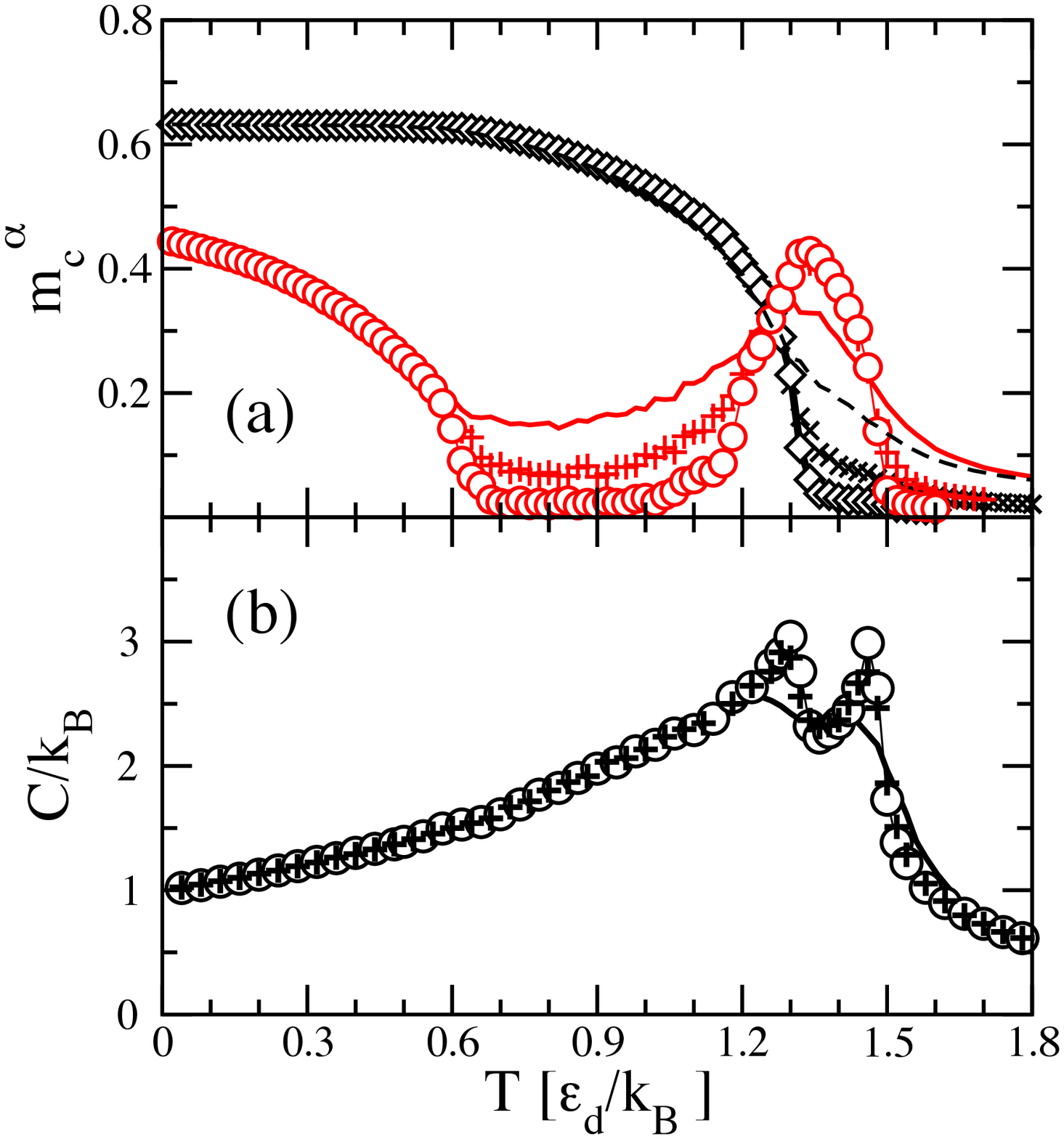}
\caption{(Color online)
(a) $m_{\mathfrak s}^\alpha$ for $\alpha = y ,z$ vs temperature
for systems of $L\times L\times L$ spins
with  $\varepsilon_d=0$, $C=4J<0$, and $D=0.8C$.
Data points for $m_{\mathfrak s}^x$ fall on top
of the data points for $m_{\mathfrak s}^y$ and
are therrefore not shown.
$m_{\mathfrak s}^y$ is represented by (red online)
$\circ$, $+$, and continuous line for $L=32$, $16$, and $8$,
respectively. $m_{\mathfrak s}^z$ is represented by (black online)
$\lozenge$, $\times$, and dashed line for $L=32$, $16$, and $8$,
respectively.
All data points come from averages of
$\mid m_{\mathfrak s}^\alpha\mid$
over $1.8\times 10^5$ MC sweeps.
(b) Same as in (a) but for $C/k_B$ vs $T$.}
\label{2SRj}
\end{figure}

It is interesting to compare the ratio $T_z/T_{xy}$ between
the two relevant temperatures for the SR transition. Some
results we have obtained from MC
simulations and from the mean field
equations above are shown in Fig. \ref{Tratio}.
How little the results seem to depend on whether
SR is brought about purely by dipolar interactions
($J=0$) or solely by exchange interactions ($\varepsilon_d=0$)
is noteworthy. This may be related to
the fact that SR is completely independent of $J$
in the ground state (see Sec. \ref{gs}). How 
close the mean field and MC data points lie
over most of the $0<D/C<1$ range
in Fig. \ref{MF} is also intriguing.

\subsection{Reverse SR}

The rather abrupt variation of $T_z/T_{xy}$
near $D/C=0.8$ (see Figs. \ref{SRph} and \ref{Tratio})
is worth exploring.

Consider first $J=0$. The behavior that obtains
as a function of $T$ for $D=-0.747C$, and,
somewhat irrelevantly, $C=-3\varepsilon_d$,
is illustrated in Figs. \ref{2SRd}(a)
and \ref{2SRd}(b). Note that
$\langle \mid m_{\mathfrak c}^z \mid\rangle$ seems to approach
zero in the $0.35\lesssim T\lesssim 0.55$ $\varepsilon_d/k_B$
range as system size increases.
Thus, four phases are visited as
$T$ varies in Fig. \ref{2SRd}(a): the paramagnetic phase,
at $0.7\varepsilon_d/k_B\lesssim T$, the $z$-collinear phase, at
$0.55 \lesssim T\lesssim 0.7$ $ \varepsilon_d/k_B$, the $xy$-canted phase,
at $0.35\lesssim T\lesssim 0.55$ $ \varepsilon_d/k_B$, and the SR phase,
at $T\lesssim 0.35$ $\varepsilon_d/k_B$. We refer to the
SR that takes place within the SR phase, where
$T\lesssim 0.35\varepsilon_d/k_B$ in Fig. \ref{2SRd}(a),
as a {\it reverse} SR, since spin orientations
then vary opposite to the way they vary
at the higher transition (near $T=0.55\varepsilon_d/k_B$).

The rather abrupt change in
$m_{\mathfrak c}^y$ and in $m_{\mathfrak c}^z$
near $T=0.7$ is
in accordance with a first order transition, as
in Landau's theory.\cite{LL} (This is so
in Landau's theory because neither symmetry
group in either of the two phases is a subgroup
of the other one. Note that the opposite
condition obtains at the other two transition
points, hence continuous transitions are
predicted therein.) 

\begin{figure}[!ht]
\includegraphics*[width=80mm]{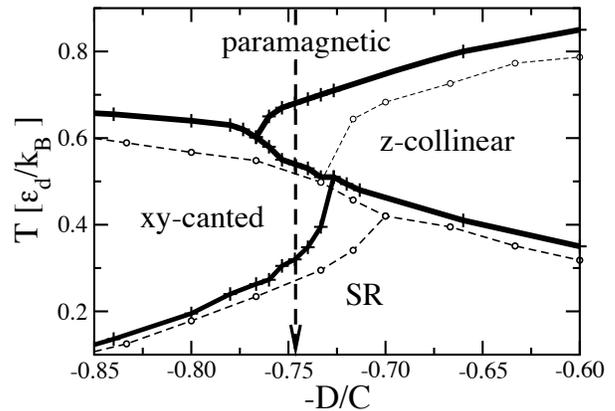}
\caption{Magnetic phase boundaries
for systems of $L\times L\times L$
spins in which $J=0$ and $C=-3\varepsilon_d$.
The full (dashed) line and $+$ ($\circ$) is for $L=16$ ($L=8$).
The vertical dashed line with an arrow
is for the cooling path from which the
data points in Fig. \ref{2SRd} follow.}
\label{revPH}
\end{figure}

Reverse SR does not seem to be an isolated phenomenon.
It also obtains for other
values of $D$, in the range $0.72\lesssim D/C\lesssim 0.77$,
and for nonzero values of the exchange constant. This
is illustrated in Figs. \ref{2SRj}(a) and \ref{2SRj}(b),
for $\varepsilon_d=0$, $D/C=0.8$, and,
less importantly, $C=4J$.
From similar plots to the one shown in Figs. \ref{2SRd}(a)
and \ref{2SRd}(b), we have obtained for various values
of $C/D$ the phase diagram shown in Fig. \ref{revPH}.

\section{conclusions}

We have studied the effect of
a fourfold anisotropy on the magnetic phases of
systems of classical
magnetic dipoles on simple cubic lattices with
dipolar and antiferromagnetic exchange interactions.
For negative anisotropy constants $D$ and $C$, and for $T=0$,
we find canted and collinear spin configurations,
as shown in Fig. \ref{pattern2}, if
$-1.34\varepsilon_d\lesssim J\leq 0$ and
$J\lesssim -1.34\varepsilon_d$, respectively. An interesting
temperature driven transition between two
ordered phases is also reported (see Fig. \ref{ree2}).

We have studied in some depth the spin reorientation phase.
Its broken symmetries are exhibited in detail.
In our model, a thermally driven continuous SR transition occurs if, in the ground state,
spins are tilted away from the crystalline axes. 
We show, by mean field theory and by MC simulations, that,
upon cooling below the paramagnetic phase,
collinear (along the easy magnetization axis)
spin configurations
obtain if $D<0$ and
$0\lesssim D/C \lesssim 0.8$, and spin reorientation
(towards the perpendicular plane) takes place at lower temperatures.
On the other hand, if if $D<0$ and $0.8\lesssim D/C<1$,
spins point perpendicularly to the easy magnetization axis
below the paramagnetic phase, and a spin reorientation
towards the easy magnetization axis takes place below a
lower transition temperature. 

Intriguingly, the ratio between the
two relevant transition
temperatures, $T_z/T_{xy}$ (see Fig. \ref{Tratio}),
seems to depend, over most of the
$0<D/C<1$ range,
neither on the strength of the exchange interaction, nor on $C$,
nor on whether the two temperatures follow from
MC simulations or from mean field thoery.

If $D<0$ and $D/C\sim 0.8$, then
(see Figs. \ref{2SRd}, \ref{2SRj}, and \ref{revPH}),
independently of the strength of the exchange interaction,
a complete spin reorientation, from the easy magnetization
axis into the perpendicular plane first takes place upon cooling
below the paramagnetic phase, followed at lower temperature by a reverse
spin reorientation, in the SR phase, towards the easy magnetization axis. 

\acknowledgments
{Financial support from Grant No. BFM2003-03919/FISI, 
from the Ministerio de Ciencia y 
Tecnolog\'{\i}a of Spain, is gratefully acknowledged.}

\end{document}